\documentclass[preprint,12pt]{elsarticle}

\usepackage{graphicx}
\usepackage{amssymb}
\usepackage{amsmath}
\newcounter{bla}

\journal{Computer Physics Communications}

\begin{document}

\begin{frontmatter}



\title{Performance Evaluation of the Zero-Multipole Summation Method in Modern Molecular Dynamics Software}


\author[ad:utokyo]{Shun Sakuraba\corref{au:sakuraba}}
\author[ad:osaka-ipr]{Ikuo Fukuda}

\cortext[au:sakuraba]{Corresponding author.\\\textit{E-mail address:} sakuraba@cbms.k.u-tokyo.ac.jp}
\address[ad:utokyo]{Graduate School of Frontier Sciences, The University of Tokyo, 5-1-5 Kashiwanoha, Kashiwa-shi, Chiba 277-8561, Japan.}
\address[ad:osaka-ipr]{Institute for Protein Research, Osaka University, 3-2 Yamadaoka, Suita-shi, Osaka 565-0871, Japan.}

\begin{abstract}
We evaluate the practical performance of the zero-multiple summation method (ZMM), a method for approximately calculating electrostatic interactions in molecular dynamics simulations. The performance of the ZMM is compared with that of the smooth particle mesh Ewald method (SPME). Even though the ZMM uses a larger cutoff distance than the SPME does, the performance of the ZMM is found to be comparable to or better than that of the SPME. In particular, the ZMM with quadrupole or octupole cancellation and no damping factor is an excellent candidate for the fast calculation of electrostatic potentials.
\end{abstract}


\end{frontmatter}

%\linenumbers

\section{Introduction}
In molecular dynamics (MD) simulations, molecules are modeled in a computer using interaction potentials between atoms, and their dynamics are simulated by numerically solving the Newtonian equations of motion. In MD simulations, a significant amount of computation time is consumed during the evaluation of electrostatic interactions. Because accurate evaluation of electrostatic interactions is vital for an accurate MD simulation, computationally efficient yet reasonably accurate ways of modeling and approximating electrostatic interactions have been of much interest in the development of MD methods. Such methodologies include the particle--particle particle--mesh method (PPPM),\cite{Eastwood1980} smooth particle-mesh Ewald method (SPME),\cite{Darden1993,Essmann1995} Gaussian-split Ewald method,\cite{Shan2005} and fast multipole method,\cite{Greengard1987,Figueirido1997} to name a few. Among them, the PPPM and SPME are arguably the most commonly used techniques in current MD simulations. Current state-of-art MD simulation packages such as AMBER,\cite{AMBER2015} CHARMM,\cite{Brooks1983,Brooks2009,Hynninen2014} GROMACS,\cite{Berendsen1995,Pall2015,Abraham2015} LAMMPS,\cite{Plimpton1995} and NAMD\cite{NAMD2005} implement either (or both) of these algorithms. Roughly speaking, the PPPM and SPME accelerate the Ewald summation by splitting the electrostatic interaction into a short-ranged real part and a long-ranged reciprocal part, where the former is computed directly and the latter is computed approximately by convolution via a three-dimensional fast Fourier transform (FFT). Unfortunately, the FFT requires several global communication steps for the ``butterfly'' exchange of intermediate values. In modern computers, communication between processors is much costlier than computations, and thus methods that can avoid FFT offer huge advantages particularly when executed in parallel.

Historically, before the development of these efficient mesh-based methods, electrostatic interactions between atoms were typically calculated by only accumulating the interactions within some cutoff distance. These methods are called as cutoff-based methods.\cite{Fukuda2012} Cutoff-based methods inherently do not require the FFT, and therefore, have a good scalability even in massively parallel environments. However, these cutoff-based methods have been mostly abandoned because of serious artifacts. For example, artifacts originated from layer structure in liquid water systems were reported when the plain-cutoff\cite{Yonetani2005,Yonetani2006,Spoel2006} or the switching force\cite{Yonetani2006} is used. Similarly, properties of several water models deviate between the reaction-field method and the Ewald-based methods.\cite{Alper1989,Hess2002} With the introduction of efficient Ewald-based methods such as the PPPM and the SPME, cutoff-based methods have been rarely used in modern molecular simulations.

Recently, several cutoff-based methods\cite{Yakub2003,Wu2005,Heinz2005,Shi2008,Fukuda2013,Fukuda2014} that can avoid such artifacts have been proposed. The zero-multipole summation method (ZMM) is one of such examples.\cite{Fukuda2013} The ZMM utilizes the neutralization of electrostatic multipoles, which is a reasonable assumption in solvents such as water-containing systems, to compute electrostatic interactions. The electrostatic energies in the ZMM have been shown to match those in the SPME to within a small deviation,\cite{Fukuda2014} and both the static and dynamic properties of bulk water computed by the ZMM have been shown to agree with the SPME,\cite{Wang2016} which makes the ZMM a promising candidate for computing approximate electrostatic potentials. The drawbacks of the ZMM are that it may require a slightly larger cutoff in real-space particle-particle interactions than does the SPME for sufficiently accurate computations. For example, \citet{Wang2016} used a cutoff distance of 1.2 nm for validation of the ZMM, whereas in the SPME the real-space cutoff distance may vary and is typically set to 1.0 nm or shorter for good performance. The ZMM's practical performance properties were thus not apparent. Although it was preliminarily reported that the zero-dipole summation method,\cite{Kamiya2013} a special case of the ZMM, runs with performance comparable to that of the SPME, there have been no extensive software engineering efforts to optimize the performance of both methods and no comparison of the ZMM using higher order multipole parameters. Especially, in the case of the SPME, it has been reported that engineering efforts are crucial for high-performance implementations.\cite{Phillips2002,NAMD2005,Hess2008a,Pronk2013,Pall2015} A rigorous comparison with the state-of-art implementation of the SPME is thus needed.

In the present article, we report the performance of the ZMM implemented in a molecular dynamics software package, GROMACS.\cite{Berendsen1995,Hess2008,Pall2015,Abraham2015} GROMACS has been implemented with a mechanism to separate real and reciprocal ranks in the SPME calculation, 2D-decomposition of 3D FFT calculation for the reciprocal calculation, SIMD-optimized non-bonded interaction calculations, and fast short-ranged Ewald kernel to accelerate the SPME calculation.\cite{Hess2008a,Pall2013,Abraham2015,Pall2015} It thus contains one of the state-of-the-art implementations of the SPME. We compare the performance of the ZMM with that of the SPME to elucidate the differences at a practical level.

In sections \ref{subsec:ZMM} and \ref{subsec:SPME}, we briefly review the formulae and the reported properties of the ZMM and the SPME, respectively. We then explain the performance optimization of the ZMM and the SPME in sections \ref{subsec:implZMM} and \ref{subsec:implSPME}, respectively. The system setup is shown in section \ref{subsec:materials}, and the accuracy measurement is reported in section \ref{subsec:accuracy-measurement}. The protocol of the performance measurement, which is the main issue of the study, is explained in section \ref{subsec:performance-measurement}. Results and discussion for the performance measurement follows in section \ref{sec:Results-and-Discussion}. Finally, we conclude the current work in section \ref{sec:Conclusion}.

\section{Methods\label{sec:Methods}}

\subsection{Zero-Multipole Summation Method\label{subsec:ZMM}}

\subsubsection{Formula\label{subsubsec:ZMM-formula}}
The ZMM assumes electric multipole neutralization in a system for fast computation of the electrostatics. More rigorously, for any particle $i$ ($1\le i\le N$ where $N$ is the number of particles), the ZMM assumes that there exists a particle set $\mathcal{M}_{i}^{\left(\ell\right)}$ such that the $\ell$th and lower order electrostatic multipoles\footnote{We refer to charges as the 0th order multipoles, dipoles as the 1st order multipoles, quadrupoles as the 2nd order multipole, and so on.} are neutralized for $\mathcal{M}_{i}^{\left(\ell\right)}\cup\left\{ i\right\} $, and the particles in $\mathcal{M}_{i}^{\left(\ell\right)}$ are within distance $r_{c}$ from particle $i$ (hereinafter this assumption is called the {\it neutralization principle}). The particles in $\mathcal{M}_{i}^{\left(\ell\right)}$ are further assumed to nearly fill the (cutoff) sphere around $i$ to a radius of $r_{c}$. Under these assumptions, the electrostatic interaction between particle $i$ and particle $j$ is suitably reformulated if their distance is less than $r_{c}$, while that is approximately replaced by a correction term if their distance is farther than $r_c$.

The potential function in the ZMM is then formed to approximate the electrostatic interaction. The obtained electrostatic interaction potential $V_{\mathrm{ZM},ij}$ between particle $i$ and $j$ is
\begin{gather}
V_{\mathrm{ZM},ij}=\frac{q_{i}q_{j}}{4\pi\epsilon_{0}}U_{\mathrm{ZM}}\left(r_{ij}\right)\\
U_{\mathrm{ZM}}\left(r\right)=
  \begin{cases}{
    \displaystyle \frac{\mathrm{erfc}\left(\alpha r\right)}{r}+\sum_{k=0}^{\ell}c_{k}r^{2k}} & \left(r\le r_{c}\right)\\
    0, & \left(r>r_{c}\right)
  \end{cases}
\end{gather}
where $\epsilon_{0}$ is the dielectric constant of the vacuum, $q_{i}$ and $q_{j}$ are the charges of atoms $i$ and $j$ respectively, $r_{ij}$ is the distance between particles $i$ and $j$, $r_{c}$ is the cutoff distance, $\alpha$ is the real-space damping factor, $\mathrm{erfc}\left(\cdot\right)$ is the complementary error function defined as
\begin{equation}
  \mathrm{erfc}\left(z\right)=1-\frac{2}{\sqrt{\pi}}\int_{0}^{z}\exp\left(-x^{2}\right)\mathrm{d}x,
\end{equation}
and $c_{k}$ are constants that depend on $\ell$, $\alpha$, and $r_{c}$. Because only the interactions within the cutoff radius $r_c$ are considered, the time complexity to run the ZMM per atom is expected to be constant, thus the total time to compute the interactions scales with $O\left(N\right)$.

\subsubsection{Accuracy\label{subsubsec:ZMM-accuracy}}

This study focuses on the computational performance of the ZMM, since its accuracy has already been clarified in previous studies. We briefly review those results in this section.
The accuracy of the ZMM is clear when applied to systems that are governed purely by electrostatic interactions. The method was used to calculate the Madelung energy of ionic crystal systems\cite{Fukuda2013}. In these pure electrostatic systems, the neutralization principle works well, and the accuracy is improved by increasing the multipole order $\ell$. The ZMM gives a high accuracy, and the absolute error rate from the exact value is, for example, of the order $10^{-6}$ at around $r_c=4a$, where $a$ is the nearest-neighbor ionic distance in NaCl crystal. This is significantly more accurate than conventional real-space methods, including the cubic minimum image convention, and methods by \citet{Harrison2006} and \citet{Gaio2009}.

The best accuracy of the ZMM in liquids is attained at $\ell =2$ or 3 with $\alpha\le 1$ nm${}^{-1}$. Ionic liquids and pure water systems are governed by both electrostatic interactions and the thermodynamic environment.\cite{Fukuda2014} In these systems, neutralized states are dominant but not always realized at a higher order multipole level due to thermal fluctuations, and so a moderate value of multipole order is efficient. The accuracies in the TIP3P system have been investigated in detail.\cite{Wang2016} Thermodynamic quantities, including the chemical potential, heat capacity, and isothermal compressibility, and dynamical quantities, including the diffusion constant and viscosity, obtained by the ZMM with $r_c=1.2$ nm agreed with the results of a highly optimized SPME computation within statistical uncertainty at a confidence level of 95\%. The dielectric properties were carefully investigated using long-time simulations\cite{Gereben2011} taking into account the correction formula\cite{Neumann1983} and system size effects.\cite{Spoel2006} The dielectric constant agreed with the SPME result, and the Kirkwood-G factor, which is very sensitive to the electrostatic treatment, also showed accurate results beyond two or three times the cutoff length and they are qualitatively excellent compared with the reaction field method with the group-based cutoff. The density deviation between the ZMM with $r_c=1.2$ nm and the optimized SPME was of the order of 0.1\%. The radial distribution error in the ZMM was less than 0.2\% with $\ell=2$, and was considerably smaller with $\ell=3$. The ZMM with $\ell=1$ has been applied to biomolecule simulations, where the systems are filled with water molecules in realistic explicit solvent studies. In explicit solvated systems of DNA\cite{Arakawa2013} and a membrane protein,\cite{Kamiya2013} the accuracies of energetic and dynamical properties for each scale level, such as molecule, residue, or nucleotide, have been confirmed by comparison with the SPME method.

\subsection{Smooth Particle Mesh Ewald Method\label{subsec:SPME}}

\subsubsection{Formula}

The SPME method approximates the total electrostatic interaction energies in a periodic boundary condition as a sum of two energies:
\begin{equation}
  V_{\mathrm{SPME}}=V_{\mathrm{real}}+V_{\mathrm{recp}},
\end{equation}
where $V_{\mathrm{real}}$ is the real-space electrostatic energy, and $V_{\mathrm{recp}}$ is the reciprocal-space electrostatic energy. The real-space electrostatic energy $V_{\mathrm{real}}$ is represented as a sum of short-ranged interaction potentials as
\begin{equation}
  V_{\mathrm{real}}=\sum_{i=1}^{N} \sum_{j=i+1}^{N} V_{\mathrm{real},ij}=\sum_{i=1}^{N} \sum_{j=i+1}^{N}\frac{q_{i}q_{j}}{4\pi\epsilon_{0}}\frac{\mathrm{erfc}\left(\beta r_{ij}\right)}{r_{ij}}.\label{eq:spme-real}
\end{equation}
Because $V_{\mathrm{real},ij}$ decreases rapidly as $r_{ij}$ becomes large for typically used value of $\beta$, the real-space electrostatic energy can be computed by simply summing interaction potentials from the atoms within a cutoff distance. As in the case of the ZMM, we denote this cutoff length as $r_{c}$.

The reciprocal-space electrostatic energy $V_{\mathrm{recp}}$ is calculated by grid-wise summation with a grid size $K_{1}\times K_{2}\times K_{3}$,
\begin{equation}
  V_{\mathrm{recp}}=\frac{1}{2}\sum_{k_{1}=0}^{K_{1}-1}\sum_{k_{2}=0}^{K_{2}-1}\sum_{k_{3}=0}^{K_{3}-1}D\left(k_{1},k_{2},k_{3}\right)\left|\mathcal{F}\left(k_{1},k_{2},k_{3}\right)\right|^{2},
\end{equation}
where $D$ is a modified Green's function in reciprocal-space and $\mathcal{F}$ is a Fourier-transformed grid charge distribution $Q$:
\begin{gather}
  \mathcal{F}\left(k_{1},k_{2},k_{3}\right)=\sum_{m_{1}=0}^{K_{1}-1}\sum_{m_{2}=0}^{K_{2}-1}\sum_{m_{3}=0}^{K_{3}-1}Q\left(m_{1,}m_{2},m_{3}\right)\nonumber\\
  \qquad\times\exp\left[-2\pi\sqrt{-1}\left(\frac{m_{1}k_{1}}{K_{1}}+\frac{m_{2}k_{2}}{K_{2}}+\frac{m_{3}k_{3}}{K_{3}}\right)\right].\label{eq:fourier-transform-q}
\end{gather}
The formulations of $D$ and $Q$ are described elsewhere.\cite{Essmann1995} 

\subsubsection{Computational Cost}
The Fourier transform (\ref{eq:fourier-transform-q}) is typically computed by a parallel three-dimensional FFT. Although the time complexity of the FFT calculation per process scales with $O\left(K_{1}K_{2}K_{3}P^{-1}\log K_{1}K_{2}K_{3}\right)$, where $P$ is the number of processors, and the total communication amount scales with $O\left(K_{1}K_{2}K_{3}P^{-1}\right)$, the number of communication messages per process grows as $O\left(P\right)$.\cite{Pekurovsky2012} Thus, even with a fixed grid size, the total amount of time needed to initiate communication increases as the number of processors increases until it eventually becomes the dominant part of the calculation for large $P$. This makes the three-dimensional FFT costly in terms of the wall-clock time to complete the calculation in a parallel environment.

To reduce the increased communication cost, many SPME implementations separate processes into two groups, namely, real-space processes and reciprocal-space processes.\cite{Phillips2002,Hess2008a} The former evaluates $V_{\mathrm{real}}$ and the corresponding forces, and sends atomic coordinates to the reciprocal-space processes. The latter computes the reciprocal space energy and forces, and sends them back to the real-space nodes. In this way, the total number of processes involved in the FFT can be decreased. For large $P$, communication between the real-space and reciprocal-space processes, as well as intra-communication within the reciprocal process group, may occupy a significant amount of time. We will illustrate this problem in section \ref{sec:Results-and-Discussion}.

\subsection{Implementation and Optimization of the ZMM\label{subsec:implZMM}}

We implemented the ZMM in GROMACS 5.0.7. Only the cases with $\ell=2,3$ were implemented for performance comparison. Since the ZMM with $\alpha=0$ can be computed efficiently because $\mathrm{erfc}\left(0\cdot r\right)/r=1/r$, we implemented the specialized version, too, as it can significantly decrease the number of floating point operations during the force and potential evaluation. To compute $\mathrm{erfc}\left(\alpha r\right)/r$ quickly, we reused a numerical trick from the SPME real-space calculation in GROMACS: 
\begin{equation}
\frac{\mathrm{erfc}\left(\alpha r\right)}{r}=\frac{1}{r}-\frac{\mathrm{erf}\left(\alpha r\right)}{r}=\frac{1}{r}-\alpha\cdot\frac{\mathrm{erf}\left(\alpha r\right)}{\alpha r},
\end{equation}
where $\mathrm{erf}\left(\cdot\right)=1-\mathrm{erfc}\left(\cdot\right)$ is the error function. The term $\mathrm{erf}\left(x\right)/x$ is then computed using a rational polynomial approximation. The corresponding force is computed similarly from another rational polynomial approximation. Using these techniques, a computational kernel for the ZMM electrostatic force with $\ell=2$, $\alpha=0$ was implemented with 33 floating point operations (flops) per interaction, and that for the ZMM with $\ell=2$, $\alpha\neq0$ was implemented with 65 flops. They are only 2 and 4 extra flops compared to the abrupt cutoff (31 flops)\footnote{The electrostatic force kernel for the abrupt cutoff in GROMACS shares the code with the one for the reaction field method. Due to this, the flops required for the abrupt cutoff kernel is increased from 29 flops to 31 flops.} and the SPME real-space calculation (61 flops), respectively. In GROMACS, highly optimized computational kernels are implemented using the SIMD (single-instruction, multiple-data) instructions of modern processors. Thus, we implemented a SIMD version of the ZMM to maximize the computational performance. The implementation is presented in the supporting information and also at the website of one of the authors.\cite{Github}

As stated in the introduction, the bulk of MD computation time is consumed by the electrostatic interaction. In the innermost loop of the MD calculation in GROMACS, interactions between the clusters of atoms are considered.\cite{Pall2013} To compute the interaction efficiently, the program must keep a list of neighboring atom clusters -- which is often called as the (Verlet) neighbor list.\cite{Verlet1967,Allen1987,Pall2013} The neighbor lists are typically constructed to store a list of atoms within the distance that is the sum of the cutoff distance and some buffer length; that is, $r_{\mathrm{nl}}=r_{c}+r_{b}$, where $r_{\mathrm{nl}}$ is the total length of the neighbor list and $r_{b}$ is the buffer length. The neighbor list is updated periodically to save computational time with a certain interval. The choice of $r_{b}$ is crucial to balance the accuracy and speed of the computation. For accurate computation, $r_{b}$ must be sufficiently large so that all atoms within the cutoff distance are included in the neighbor list until the next list update. Meanwhile, large $r_{b}$ leads to a performance degradation due to the increased calculation cost. In GROMACS 5, the buffer length $r_{b}$ is determined so that the expected energy drift in the simulation does not exceed the limit.\cite[Section 3.4.2]{ManualGromacs5} This scheme is often critical to attain a good performance, and we also utilized the scheme for the ZMM. The details are described in \ref{sec:verlet-buffer}.

We also optimized the neighbor list update interval. The performance of the ZMM is measured with a neighbor list update interval of 10, 20, or 40 steps, and the fastest parameter was selected. Although GROMACS supports hybrid parallelism with both OpenMP and MPI, preliminary tests showed that flat MPI execution is better than execution with OpenMP under most conditions. We therefore did not use OpenMP parallelization for the ZMM.

The following ZMM parameters were chosen for the comparison. (1) $\ell=2,$ $r_{c}=1.2$ nm and $\alpha=0.0$ nm${}^{-1}$; (2) $\ell=3$, $r_{c}=1.2$ nm and $\alpha=0.0$ nm${}^{-1}$; (3) $\ell=2$, $r_{c}=1.2$ nm and $\alpha=1.0$ nm${}^{-1}$; (4) $\ell=2$, $r_{c}=1.4$ nm and $\alpha=0.0$ nm$^{-1}$. In the present research, we named these settings  (1) ZQuad, (2) ZOct, (3) ZQuad-$\alpha$1, and (4) ZQuad-1.4. ZQuad and ZOct represent the ZMM runs with the most-typical parameters. ZQuad-$\alpha$1 and ZQuad-1.4 were tested to show the performance difference with a non-zero dumping factor and with an increased cutoff length. We also benchmarked conditions with $\ell=2,$ $r_c=1.2$ nm and different $\alpha$ values, $\alpha=0.5$ nm${}^{-1}$ and $\alpha=1.5$ nm${}^{-1}$ in a preliminary analysis for the case of 10,000 TIP3P molecules (see section \ref{subsec:materials}). The results showed that the performance differences in comparison with condition (3) were small (within $\pm$3\%), and thus these conditions were not examined in larger systems. 

\subsection{Optimization of the SPME\label{subsec:implSPME}}

The SPME performance was optimized by tweaking the SPME run parameters\cite{Kutzner2015} for careful comparison of the ZMM and the SPME. Fundamental SPME parameters were taken from the default values in GROMACS. The minimum grid distance is 0.12 nm and the SPME spline order was set to 4. The following parameters were considered in the optimization: (1) the SPME was performed with separate reciprocal processes or without separate reciprocal processes (i.e., the real and reciprocal calculations were performed on all processes), (2) various numbers of real-reciprocal process number ratios (the number of reciprocal processes were searched within 15\%--50\% of the real-space cores), (3) runs were performed with and without OpenMP parallelization with two threads, (4) the compensation between the real-space cutoff distance and the reciprocal grid size (up to 20\% increase in the cutoff distance and the grid size) was considered, and (5) the neighbor list was updated every 10, 20, or 40 steps. Some of the parameters were optimized using a tool \texttt{gmx tune\_pme} in GROMACS. The fastest combination of (1)-(4) was first searched, and then the best neighbor list update interval (5) was searched to find the optimal condition for the SPME. We set the minimum cutoff distance to 1.0 nm, because the Lennard-Jones interaction cutoff distance must be equal to the electrostatic cutoff for the optimal performance in GROMACS 5, and lowering the cutoff below 1.0 nm showed non-negligible stability differences for the case of proteins.\cite{Piana2012} We call the SPME calculation with the optimal parameter set SPME-1.0. We also tried the Lennard-Jones Particle Mesh Ewald (LJ-PME)\cite{Essmann1995} to have a shorter real-space cutoff, but preliminary analysis showed that the combination of the SPME and the LJ-PME with 0.9 nm cutoff was slower than SPME-1.0 (data not shown), and thus we did not include the LJ-PME in the comparison. In addition to the SPME-1.0, we also performed an SPME simulation with 1.2 nm minimum cutoff for comparison with the ZMM, which we call SPME-1.2.

\subsection{Materials\label{subsec:materials}}

We investigated water systems in detail for fundamental considerations and also protein systems in addition.

\paragraph{Water systems} Since water is necessary for biomolecular simulations and since the accuracy of the ZMM in TIP3P water systems was revealed in Ref. \cite{Wang2016}, detailed investigation of the computational performance of the ZMM in the TIP3P system is a current principal interest. We prepared four systems with TIP3P\cite{Jorgensen1983} water boxes containing 10,000, 20,000, 40,000, and 80,000 molecules, respectively, meaning 30,000, 60,000, 120,000, and 240,000 atom systems, respectively.

\paragraph{Protein systems} As an application to more realistic systems, we benchmarked the ZMM calculation of two systems, Trp-cage and $\gamma\delta$TCR proteins. The Trp-cage is a mini-protein consisting of 10-residue amino acids, consisting of 15,596 atoms including water and ions. The $\gamma\delta$TCR is a 206-residue protein dimer (412 residues in total), solvated in water and ions consisting 93,794 atoms. The trp-cage was modeled as a triclinic periodic boundary system while the $\gamma\delta$TCR as a rhombic dodecahedron periodic boundary system. 

\subsection{Accuracy measurement\label{subsec:accuracy-measurement}}

Prior to the performance measurement, we have carefully investigated the accuracy of the ZMM for the mainly concerned system, TIP3P water. Although the accuracy of the TIP3P system was revealed in Ref. \cite{Wang2016}, we verified the accuracies for the specific systems, including the ones with over $20,000$ molecules, prepared in the current performance study. First we measured the relative differences in both electrostatic energies and densities between the SPME and ZMM to confirm the correctness of the implementation. The mean electrostatic energy $\bar{E}_{\text{el,}\mathsf{M}}$ was measured as the mean of $1000$ total electrostatic energies of method \textsf{M}, $E_{\text{el,}\mathsf{M}}$, which were obtained from a $1$ ns NPT MD run at $310$ K and $1$ atm using method \textsf{M} after an equilibration. The relative difference in electrostatic energies is defined as $\left\vert \left(  \bar{E}_{\text{el,ZMM}}-\bar{E}_{\text{el,SPME-1.2}}\right) / \bar{E}_{\text{el,SPME-1.2}}\right\vert $. The relative difference in the densities was obtained similarly. We confirmed that the relative differences of electrostatic energies and densities between the SPME-1.2 and the ZMM are less than $4.1\times 10^{-4}$ and $1.1\times 10^{-3}$, respectively, in all three ZMM parameters with $r_c=1.2$ nm (ZQuad, ZOct, ZQuad-$\alpha1$) and all system sizes; these results are consistent to the reported deviations in previous literatures\cite{Fukuda2014,Wang2016} and in section \ref{subsubsec:ZMM-accuracy}. The standard deviations of the electrostatic energies with the SPME and the ZMM are presented in Table S1. In all the three ZMM parameters and the four system sizes, differences in $\bar{E}_{\text{el,}\mathsf{M}}$ between the SPME and the ZMM were within the margin of the standard deviations.

We further studied the accuracy about the electrostatic energy via the differences in the energy between two molecule conformations. This test can effectively capture the deviations among different methods, because it employs the pair of conformations so that the data are larger than that in the ordinary energy comparison employing every single conformation.\cite{Fennell2006} After an equilibration, NPT simulations at $310$ K and $1$ atm for the TIP3P water systems with different sizes were performed for $1$ ns with SPME-1.2 (this protocol is the same as that in the performance measurement described in section \ref{subsec:performance-measurement}, except the simulation length), and conformation snapshots in the trajectories were recorded every $10$ ps. The electrostatic energies of these snapshots were then evaluated by SPME-1.2, ZQuad, ZQuad-$\alpha 1$, and ZOct. For all possible combinations of two snapshots, the energy differences were computed by SPME-1.2 and the ZMMs independently, which made $9900\times 3$ total data.

Figure S1 shows the correlation plots between SPME-1.2 and the individual ZMM for the $10,000$ TIP3P molecule system. The linear regression between the two methods shows an excellent agreement, indicated by $R^{2}\ge 0.9997$ for all ZMMs. This result was similar for the larger systems having $20000$, $40000$, $80000$ TIP3P molecules, as shown by the regression coefficients and correlation coefficients that are also close to unity (Table S2). These results indicate that the all ZMMs with the different parameters were equally accurate for the water systems used in the performance measurement, which is the main issue of this paper, as detailed below. For further investigation of accuracy for other physical quantities and for other materials, readers can consult the literature stated in section \ref{subsubsec:ZMM-accuracy}.

\subsection{Performance Measurement\label{subsec:performance-measurement}}

The performance of the ZMM was measured and compared with that of the SPME in the water systems and protein systems. The benchmark was taken on the Cray XC30 supercomputer at Kyoto University. Each node was composed of two Xeon E5-2695v3 CPUs having 14$\times$2 physical processor cores, and nodes were connected by the Aries interconnect. Benchmarks were taken with 1, 2, 4, 8, 12, and 16 nodes (28, 56, 112, 224, 336, and 448 cores respectively). Throughout the simulations, hyperthreading was not used. The program was compiled with GCC 4.9.2 and FFTW-3.3.3\cite{fftw} was used as the FFT library. The simulation conditions were set up as follows: the time step for the simulation was set to 2.5 fs and hydrogens connected to heavy atoms were constrained to fixed lengths by LINCS.\cite{Hess1997,Hess2008} The SETTLE algorithm was used to simulate rigid TIP3P water molecules efficiently.\cite{Miyamoto1992} The systems were equilibrated under SPME-1.2 conditions for 2 ns NPT runs. Simulations were performed with the NPT ensemble using the Parrinello-Rahman barostat and Langevin dynamics. The system was kept at 310 K and 1 atm. The simulation was run for 200 ps to measure the performance.

% does ``for the performance'' really needed??
\section{Results and Discussion for the Performance\label{sec:Results-and-Discussion}}

\begin{figure}[tbp]
\includegraphics[width=1\textwidth]{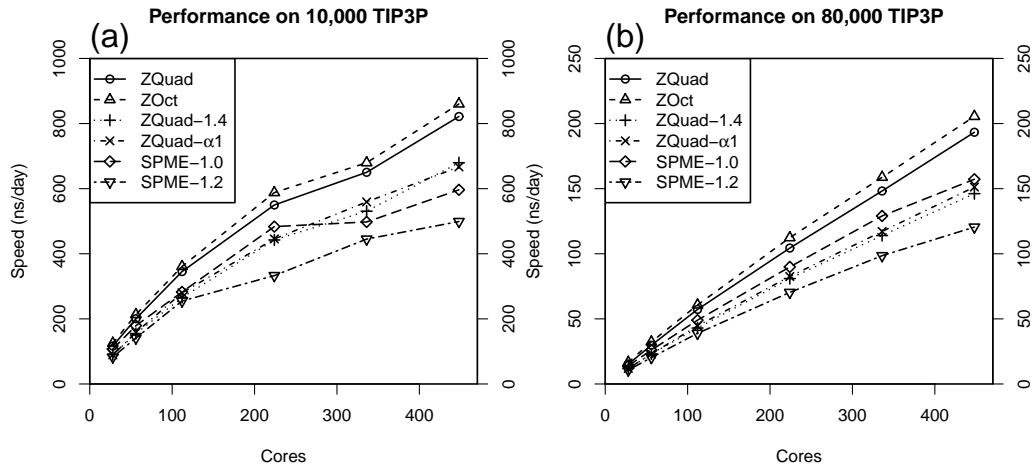}

\caption{Performance of MD simulations in GROMACS using the ZMM and the SPME with various parameters. The abscissa shows the number of physical CPU cores. The ordinate shows the speed of the MD simulation as measured in ns/day (higher values are better). The simulation setup and parameters are described in the main text.\label{fig:performance-per-cores}}
\end{figure}

\begin{figure}[tbp]
\includegraphics[width=1\textwidth]{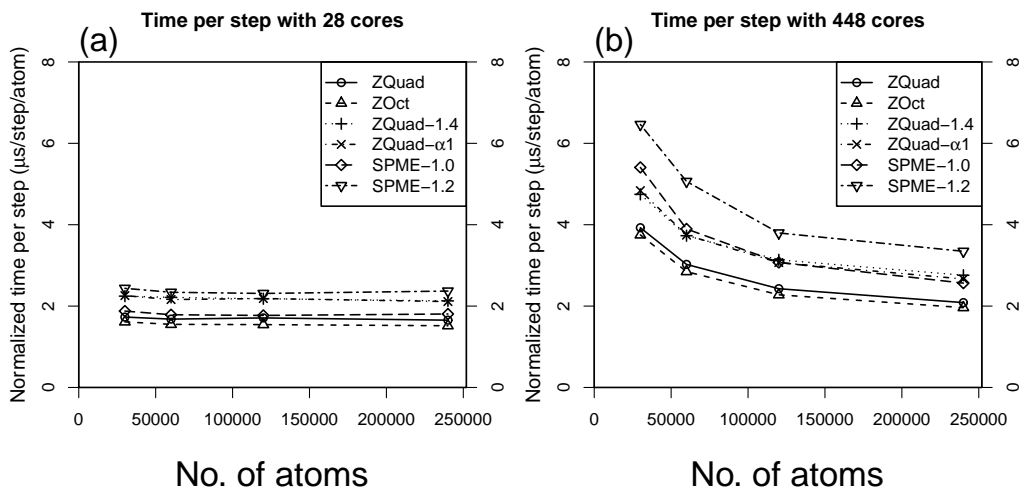}
\caption{Average time to process a single atom in a single CPU core. (a) Measured from runs with 28 CPU cores. (b) Measured from runs with 448 cores. The abscissa shows the number of simulated atoms in the system, and the ordinate shows the time required to process a single step of the simulation divided by the number of atoms per CPU core. \label{fig:performance-per-size}}
\end{figure}

\begin{figure}[tbp]
\includegraphics[width=1\textwidth]{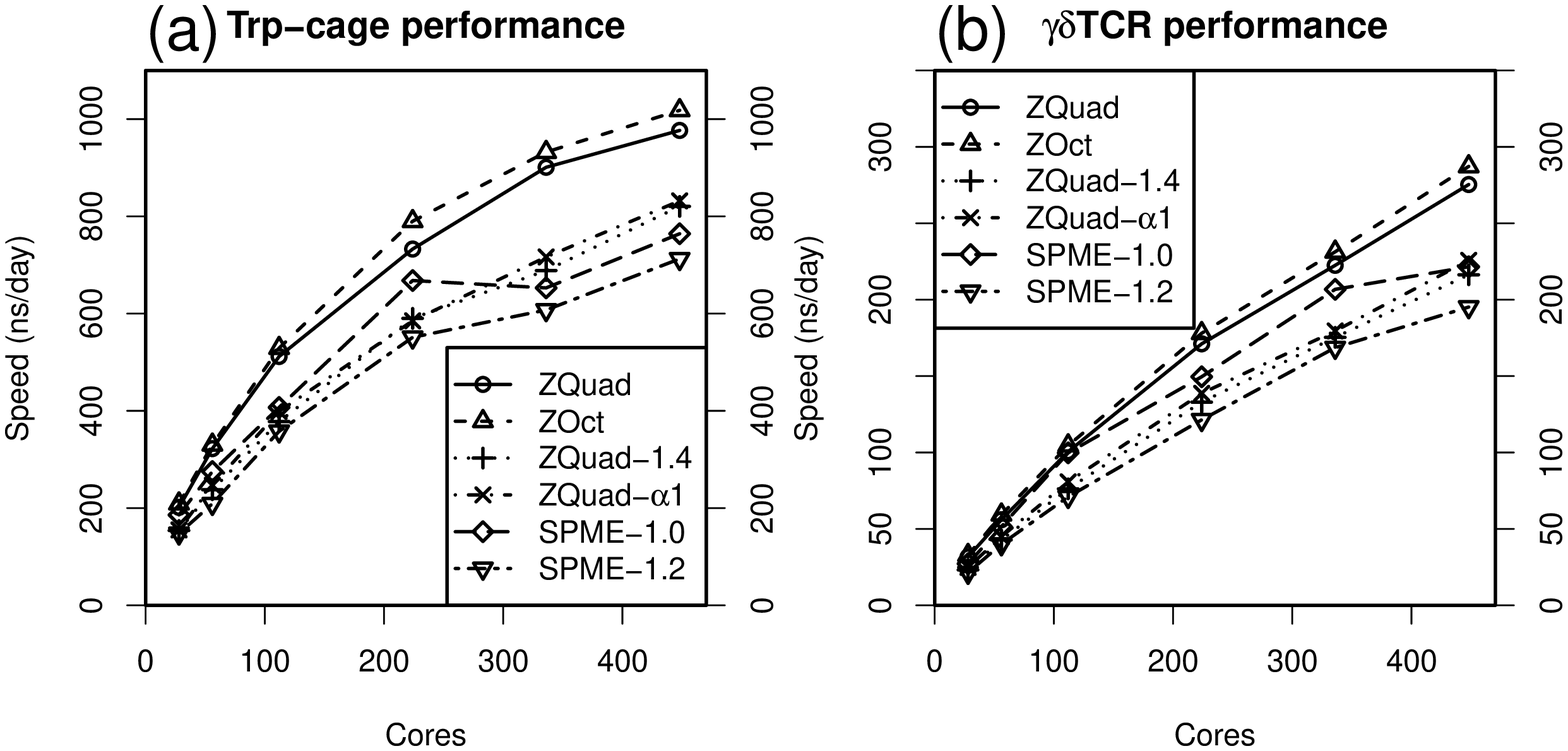}
\caption{Performance of MD simulations of proteins using the ZMM and the SPME. The abscissa shows the number of physical CPU cores and the ordinate shows the speed of the MD simulation. (a) Trp-cage mini-protein consisting of 15,596 atoms. (b) $\gamma\delta$TCR protein consisting of 93,794 atoms.\label{fig:performance-typical}}
\end{figure}

Figure \ref{fig:performance-per-cores} shows a comparison of the performance of the ZMM and SPME. ZQuad outperformed SPME-1.2 under all conditions. This is an expected result as the ZMM with $\alpha=0$ nm$^{-1}$ has a simpler form of the function, and thus requires a lower number of floating point operations. Furthermore, reciprocal convolution is not necessary in the ZMM, which also contributes favorably to the performance of the ZMM. ZQuad also showed comparable or better performance compared to SPME-1.0. This is because the performance loss from the longer cutoff length in ZQuad was compensated by the lack of reciprocal convolution part. The performance difference is more prominent for a larger number of cores. This is possibly due to the increased communication time in the SPME. Both in the 3D-FFT calculation and in the data exchange between the real and reciprocal processes, the relative cost for communication increases as the number of cores grows. This is reflected by the fact that in the SPME-1.0 simulation run with 30,000 atoms and 448 cores, 5.3\% of the wall-clock time was consumed waiting for the real-space processes to receive the force information from the reciprocal processes, whereas only 0.7\% of the wall-clock time was consumed by the SPME-1.0 run on 112 cores with the same system. 

The performance advantage of the ZMM was even more prominent for $\ell=3$, as indicated by the ZOct in Fig. \ref{fig:performance-per-cores}. ZOct exhibits a smoother potential function near $r=r_{c}$ than ZQuad, and this leads to a smaller energy drift by the neighbor list (see \ref{sec:verlet-buffer}). Consequently, a shorter neighbor list cutoff buffer length is required, and this is the main source of the performance increase. For example, in the 30,000 atoms case with a neighbor list update interval of 20 steps, the neighbor list distance $r_{\mathrm{nl}}$ of ZOct was 1.294 nm, while that of ZQuad was 1.350 nm. As a result, ZOct exhibited an increased speed over ZQuad of from 4\% to 11\% due to the lower number of interactions to compute.

ZQuad-$\alpha1$ was noticeably slower than ZQuad. This is because the computation of $\mathrm{erfc}\left(\alpha r\right)/r$ is costly. The performance of ZQuad-$\alpha$1 was comparable with or slower than that of SPME-1.0 except for cases with 336 and 448 cores in the 10,000 TIP3P system. ZQuad-1.4 also exhibited similar performance to ZQuad-$\alpha1$. From these results, it is not worth choosing these electrostatic potentials in terms of computation speed compared to SPME-1.0, at least for $\le224$ cores under the present benchmark conditions. When the number of cores is larger, these ZMM settings may become beneficial due to the lack of a reciprocal part, as is the case in the 30,000 atoms case. Overall, ZQuad-$\alpha$1 and ZQuad-1.4 exhibited a performance decrease of 12\% to 24\% and 17\% to 25\%, respectively, compared with ZQuad. 

Judging from these results and the accuracy of the ZMM presented in the ``Methods'', the ZMM with $\ell=2,3$, $r_{c}=1.2$ nm, and $\alpha=0$ nm$^{-1}$ is promising for fast simulations, particularly in medium- to high-level parallel environments.

We further analyzed the time required to process an atom to investigate the difference between the ZMM and the SPME. Figure \ref{fig:performance-per-size} shows the time to process an atom in systems of different sizes. The average time $s$ to process a single atom in a single CPU core is given by the following equation.
\begin{equation}
  s = \frac{(\text{Time to run a single MD step})}{(\text{\# of atoms in the system}) / (\text{\# of CPU cores})}.
\end{equation}
For all methods, the time per step per atom using 28 cores shown in Fig. \ref{fig:performance-per-size} (a) remains almost constant irrespective of the number of atoms. This is consistent with the time complexity of the ZMM, $O\left(N\right)$, as explained in section \ref{subsec:ZMM}. For the SPME, while the time to process each atom is expected to scale as $O\left(N\log N\right)$,\cite{Essmann1995} this was not observed in the present results. With the present range of $N \le 240,000$ and 28 CPU cores, the reciprocal calculation part was only ${}<20$\% of the total calculation time, and was thus dominated by other calculations in the MD simulation. In the case of using 448 cores in Fig. \ref{fig:performance-per-size} (b), the time to process each atom decreased as the number of atoms increased. The increase in the number of atoms under a fixed number of cores indicates a decrease in inter-node communications relative to intra-node communications, if the latter cost is not significant. This is the case for the 448 cores. However, this is not the case for the case of 28 cores shown in Fig. \ref{fig:performance-per-size} (a), where all communication is intra-node such that the time to process each atom is almost constant. This is also related to the fact that the average time when using 448 cores is longer than that when using 28 cores. In fact, the inter-node communication between processes for 448 cores is much slower than intra-node communications for 28 cores.

The increase in the average time for many cores is relatively significant for the SPME. In particular, the SPME-1.2 for 448 cores takes considerably longer time per step than the ZMM does. This also confirms the increased communication time for the reciprocal-space part in parallel environments. This point can be clarified by a direct comparison between SPME-1.2 and ZQuad-$\alpha$, since the real-space parts of SPME-1.2 and ZQuad-$\alpha$1 have comparable computational costs. For example, in the case of 30,000 atoms and 28 cores, SPME-1.2 required 2.44 $\mu$s to process an atom, only 8\% slower than ZQuad-$\alpha$1 which took 2.26 $\mu$s on average. In contrast, in the case of 448 cores, the difference between the two increased to 34\% (6.46 $\mu$s and 4.84 $\mu$s, respectively). 

Figure \ref{fig:performance-per-size} also indicates, as was observed in Fig. \ref{fig:performance-per-cores}, that ZOct exhibited the highest performance among the methods considered, and that ZQuad-1.4 exhibited performance similar to that of ZQuad-$\alpha$1.

Figure \ref{fig:performance-typical} shows the performance of the ZMM and the SPME for the Trp-cage and $\gamma\delta$TCR protein systems. The performance characteristics of the ZMM compared to the SPME in the water systems are retained even with realistic setups, including inhomogeneous systems (both Figs. \ref{fig:performance-typical} (a) and \ref{fig:performance-typical} (b)) and the rhombic dodecahedral periodic boundary conditions (Fig. \ref{fig:performance-typical} (b)). In particular, the overall behavior and relative differences among the ZMM runs with different parameters and the SPME are similar for the 30,000-atom TIP3P system (Fig. \ref{fig:performance-per-cores} (a)) and the 15,000-atom Trp-cage system (Fig. \ref{fig:performance-typical} (a)), and are also similar for the 240,000-atom TIP3P system (Fig. \ref{fig:performance-per-cores} (b)) and the 100,000-atom $\gamma\delta$TCR system (Fig. \ref{fig:performance-typical} (b)). These results imply that the performance of the ZMM is mainly governed by the system size and is unrelated to the system details that are typical to bimolecular simulations. We thus conclude that the ZMM is also beneficial in terms of performance for real-world applications, including protein MD simulations.

As in the case of the SPME, there are several important parameters in the ZMM that are tunable for better performance, while they are out of the scope of the current study. First, testing the accuracy with respect to the neighbor list cutoff distance may be beneficial for pursuing further performance. As we have shown in the comparison between the ZQuad and ZOct methods, the performance of the ZMM was strongly affected by the neighbor list distance. The total drift energy is currently estimated from the sum of absolute values of Eq. (\ref{eq:neighbor-estimation}) to give a conservative evaluation. In contrast to the Lennard-Jones interactions, electrostatic interactions may cancel each other out, and the drift energy may thus be overestimated. The development of a better drift energy estimation method may contribute to better performance of both the ZMM and the SPME.

Second, the ZMM with decreased cutoff lengths below $1.2$ nm, which were not examined in the current study, becomes an interesting topic for exploration. In this regard, the trade-off between the accuracy and the speed should be evaluated. The importance of such evaluations has been revealed in recent studies. In fact, Nozawa \textit{et al}.\cite{Nozawa2015b} discussed deviations between the Ewald method and a family of periodic reaction field methods near the phase transition point in a low-charge-density polymer system. Also for the Ewald method, Hub \textit{et al}.\cite{Hub2014} investigated artifacts that appear in a potential of mean force calculation for a non-zero net charge membrane/water system under an application of the Ewald method. These studies suggest that the accuracy investigations also for the ZMM under various conditions including that away from equilibrium should be important.

\section{Conclusion\label{sec:Conclusion}}

We implemented the ZMM, a method for calculating electrostatic interactions, in the GROMACS molecular dynamics software package and compared its performance with the SPME. We showed that even though a larger cutoff length is required in the ZMM than in the SPME, the ZMM provides performance comparable to or better than that of the SPME. In the case of medium to high parallelism in particular, the ZMM is a strong competitive alternate to the SPME in terms of performance.

\bibliographystyle{elsarticle-harv}
\bibliography{impl-paper-ref}

%%%%%%%%%%%%%%%%%%%%%%%%%%%%%%%%%%%%%%%%%%%%%%%%%%%%%%%%%%%%%%%%%%%%%
\section{Acknowledgement}

The authors would like to thank Dr. Kota Kasahara for providing the Trp-cage structure. Some of the computations were performed at the Supercomputer Center in Kyoto University, Kyoto, Japan, through the HPCI research project (project ID: hp150065, hp160165), and Research Center for Computational Science, Okazaki, Japan. This research was supported by a Grant-in-Aid for Scientific Research on Innovative Areas (No. JP221S0002) from the Japanese Ministry of Education, Culture, Sports, Science and Technology, and by the ``Development of core technologies for innovative drug development based upon IT'' from Japan Agency for Medical Research and Development (AMED).

\appendix

\section{Verlet Buffer Distances}\label{sec:verlet-buffer}
  
Given a potential function $V_{ij}\left(r\right)$, the expected drift $\left\langle \Delta V_{ij}\right\rangle$, caused by atom $j$ violating the neighbor list of atom $i$, is approximated as follows.\cite{ManualGromacs5}
\begin{align}
\left\langle \Delta V_{ij}\right\rangle & =\int_{0}^{r_{c}}\mathrm{d}r_{t} \int_{r_{\mathrm{nl}}}^{\infty}\mathrm{d}r_{0} 4\pi r_{0}^{2}\rho_{j}V_{ij}\left(r_{t}\right)G\left(\frac{r_{t}-r_{0}}{\sigma_{ij}}\right) \label{eq:neighbor-estimation-exact}\\
& \approx 4\pi\left(r_{\mathrm{nl}}+\sigma_{ij}\right)^{2}\rho_{j}\int_{-\infty}^{r_{c}}\mathrm{d}r_{t}\int_{r_{\mathrm{nl}}}^{\infty}\mathrm{d}r_{0} \left[\sum_{k=1}^{n_{k}}V_{ij}^{\left(k\right)}\left(r_{c}\right)\cdot\frac{1}{k!}\left(r_{t}-r_{c}\right)^{k}\right]\nonumber\\
&\qquad\qquad\qquad\qquad \times G\left(\frac{r_{t}-r_{0}}{\sigma_{ij}}\right),\label{eq:neighbor-estimation}
\end{align}
where $\rho_{j}$ is the number density of the atom $j$, $\sigma_{ij}$ is the standard deviation of the distance between freely moving (i.e., gas phase) particles $i$ and $j$ between updates, $V_{ij}^{\left(k\right)}\left(r_{c}\right)$ is the $k$th derivative of the potential function $V_{ij}\left(r\right)$ at $r=r_{c}$, $n_{k}$ is the order of the Taylor expansion, and $G\left(\cdot\right)$ is the normal distribution with zero mean and unit variance. From Eq. (\ref{eq:neighbor-estimation-exact}) to Eq. (\ref{eq:neighbor-estimation}), $V^{(0)}_{ij}(r_c) = 0$ is assumed and the lower bound of $r_t$ was changed to negative infinity. We used Eq. (\ref{eq:neighbor-estimation}) to determine an appropriate $r_{\mathrm{nl}}$ distance for both the ZMM and the SPME so that the expected sum of drifts from both the electrostatic and the Lennard-Jones potentials do not exceed the maximum allowed $\left|\left\langle \Delta V_{ij} \right\rangle\right|$. The value of maximum $\left|\left\langle \Delta V_{ij} \right\rangle\right|$ was set to 0.005 kJ / mol / ps / atom in this research (the default value in GROMACS 5). In GROMACS 5, the order of Taylor expansion $n_{k}$ is only up to $n_{k}=2$ for electrostatic interactions. Because the ZMM electrostatic potential satisfies $V_{ij}^{\left(k\right)}\left(r_{c}\right)=0$ for $k\le\ell$, using $n_{k}=2$ underestimates the absolute value of drift to be zero with any buffer size for $\ell \ge 2$. We thus increased the order of the Taylor expansion for the ZMM to prevent this underestimation. The third order expansion $n_{k}=3$ was used for $\ell=2$, and $n_{k}=4$ was used for $\ell=3$.

%%%%%%%%%%%%%%%%%%%%%%%%%%%%%%%%%%%%%%%%%%%%%%%%%%%%%%%%%%%%%%%%%%%%%
%\section{Supporting Information}
%The following files are available free of charge.
%\begin{itemize}
%  \item patches.zip: Patches to implement the ZMM in GROMACS 5.0.7.
%\end{itemize}

\end{document}